\documentclass[12pt]{article}
\usepackage{graphicx,epsfig,color}
\renewcommand{\theequation}{\arabic{section}.\arabic{equation}}
\newcommand{\bc}{\begin{center}}
\newcommand{\ec}{\end{center}}
\newcommand{\be}{\begin{equation}}
\newcommand{\ee}{\end{equation}}
\newcommand{\bea}{\begin{eqnarray}}
\newcommand{\eea}{\end{eqnarray}}
\newcommand{\ba}{\begin{array}}
\newcommand{\ea}{\end{array}}
\newcommand{\lb}{\label}
\newcommand{\rf}{\ref}
\newcommand{\bfg}{\begin{figure}[htbp]}
\newcommand{\efg}{\end{figure}}
\newcommand{\pr}{Phys. Rev. }
\newcommand{\np}{Nucl. Phys. }
\newcommand{\prl}{Phys. Rev. Lett. }
\newcommand{\prp}{Phys. Rep. }

\newcommand{\pl}{Phys. Lett. }

\newcommand{\jp}{J. Phys. }
\newcommand{\rmp}{Rev. Mod. Phys. }
\newcommand{\nc}{Nuovo Cimento }
\newcommand{\ptp}{Prog. Theor. Phys. }

\begin{document}

\vspace*{1. cm}
\bc
{\large \textbf{Gauge invariant bound state equations\\
\vspace{0.25 cm}
for quark-antiquark systems in QCD}}\\
\vspace{1 cm}
H. Sazdjian\\
\textit{Institut de Physique Nucl\'eaire, CNRS/IN2P3,\\
Universit\'e Paris-Sud, F-91405 Orsay, France\\
E-mail: sazdjian@ipno.in2p3.fr}
\ec
\par
\renewcommand{\thefootnote}{\fnsymbol{footnote}}
\vspace{0.75 cm}

\bc
{\large Abstract}
\ec
\par
Using gauge invariant quark Green's functions, defined with 
path-ordered gluon field phase factors along polygonal lines,
and functional relations among them, two compatible bound state 
equations of the Dirac type are established for quark-antiquark 
systems, each relative to the quark or to the antiquark of the
system. The kernels of the bound state equations are defined 
through a series of Wilson loop averages along closed polygonal 
contours and their functional derivatives on them. A sufficient 
criterion for spontaneous chiral symmetry breaking is derived, 
relating the Goldstone boson wave function in the zero total 
momentum limit with the scalar part of the gauge invariant quark 
two-point Green's function. 
\par
\vspace{0.5 cm}
\par

\newpage
\section{Introduction} \lb{s1}
\setcounter{equation}{0}

Gauge invariant quark Green's functions (GIQGF) \cite{m,nm}, together 
with Wilson loops \cite{w,p,mm,mgd,mk,dv,bnsg}, represent the natural 
tools for the investigation of the properties of observable 
quantities in QCD \cite{bw,ef,k,bmpbbp,bcpbmp,sdks,js}. Approaches using 
these ingredients meet, however, difficulties arising mainly from the 
nonlocal structure of the GIQGFs. For this reason, calculations of physical 
quantities like scattering amplitudes, bound state energies, form factors, 
have usually been carried out up to now with the more familiar formalism 
of ordinary, gauge variant, Green's functions using particular gauges.
Nevertheless, a gauge invariant formalism would bring several advantages
that are worth considering. First, one expects to find in the quantities
under consideration an infrared safe behavior, free of artificial 
singularities and divergences. This is also true for the spectral 
functions underlying the gauge invariant Green's functions. Second, 
Wilson loops, when saturated for instance by minimal surfaces, allow for 
a systematic study of the confining properties of the theory. Third, the 
resolution of bound state problems provides the knowledge of gauge 
invariant bound state wave functions which are particularly useful for 
the calculation of matrix elements of operators involving path-ordered
phase factors.
\par
In this respect, to optimize the methods of investigation based on GIQGFs,
an approach was undertaken by the present author with the aim of obtaining
integrodifferential equations that the latter would satisfy, in
a parallel way as for the Dyson-Schwinger equations in the case of
ordinary Green's functions \cite{d,schw,alklvs,fisc}. This was possible 
in the case of two-point GIQGFs (2PGIQGF) defined with a path-ordered gluon
field phase factor along polygonal lines between the quark fields \cite{s1}.
The 2PGIQGFs can then be classified according to the number of segments
their phase factor line contains. Functional relations are then 
obtained among these 2PGIQFs, which, together with the equations of
motion relative to the quark fields, lead to an integro\-differential 
equation satisfied by the 2PGIQGF defined with one straight line segment 
for the phase factor. The kernel of this equation involves, with increasing 
complexity, a series of Wilson loop averages along polygonal contours 
and their functional derivatives.
\par 
As a first step for the resolution of the above equation and the 
determination of the most important piece of the kernel, the case of
two-dimensional QCD in the large-$N_c$ limit \cite{th1} was considered. 
The equation could then be solved exactly and analytically, displaying 
the main features of the spectral properties of the quark fields 
\cite{s2}: The quarks contribute to the 2PGIQGF like physical 
particles with positive energies, respecting the causality property; the 
singularities of the GIQGF are located on the positive real axis of the 
momentum squared variable (timelike region) and are represented by
an infinite series of branch cuts. Lehmann's positivity conditions of
the spectral functions \cite{lh} are also satisfied. Although results 
obtained in two-dimensional theories cannot straight\-for\-wardly be 
transposed into four dimensions, they underline here the following two 
features: (i) the equation obtained for the 2PGIQGF is neither empty, nor 
unresolvable. With plausible assumptions about the properties of Wilson 
loops, it might also be analyzed in four dimensions. In particular, the 
only part of the kernel that survives in two dimensions is precisely the
simplest Wilson loop, corresponding to a triangular contour.
(ii) The resolution of the equation has provided new results, 
not known previously from more conventional approaches.  
\par
The aim of the present paper is to enlarge the scope of investigations
of the GIQGFs by also including in it the four-point GIQGFs (4PGIQGF),
which allows us to study the bound state problem of quark-antiquark
systems with a gauge invariant formalism. This is done by using again
the functional relationships between GIQGFs with different numbers of
segments on their polygonal lines. One then ends up
with two bound state equations of the Dirac type, each relative to 
the quark or to the antiquark of the system. The two equations are
compatible among themselves due to the validity of the Bianchi identities
satisfied by the gluon fields. The kernels of these equations involve, 
as in the 2PGIQGF case, series of Wilson loop averages along polygonal 
contours and their derivatives, as well as the 2PGIQGFs of each quark 
field. A functional relationship is established between these kernels 
and those of the 2PGIQGFs. It is assumed, for later investigations, that 
the bound state wave functions satisfy usual spectral properties, based 
on the positivity of the energies of the quark and gluon fields and on 
causality, leading here to a generalization of the 
Deser-Gilbert-Sudarshan representation \cite{dgs}.
\par
The bound state equations thus obtained allow us to further investigate  
the question of spontaneous chiral symmetry breaking. It is shown that
in the chiral limit (massless quarks), the bound state equations possess
a massless solution with zero total momentum if the 2PGIQGF possesses
a normalizable nonvanishing scalar part in this limit. This result 
is the analogue of the one established by Baker, Johnson and Lee \cite{bjl}
for the Bethe-Salpeter equation \cite{spb,gml,nk} and provides,
prior to the resolution of the bound state equations, a sufficient
criterion for chiral symmetry breaking in the case of bound states made 
of quarks and antiquarks with different flavors.   
\par
The plan of the paper is the following. In Sec. \rf{s2}, properties of
the 2PGIQGFs are summarized. In Sec. \rf{s3}, 4PGIQGFs are introduced
and their properties are displayed. In Sec. \rf{s4}, the bound 
state equations are established. Section \rf{s5} deals with the question 
of the spectral representation of the wave functions. In Sec. \rf{s6}, 
a criterion for chiral symmetry breaking is derived. Summary and 
concluding remarks follow in Sec. \rf{s7}. An Appendix is devoted to 
the normalization condition of the wave functions.
\par 

\section{Two-point Green's functions} \lb{s2}
\setcounter{equation}{0}

We summarize in this section the main results obtained for the 2PGIQGFs
in Ref. \cite{s1}. We shall mainly be interested in path-ordered phase 
factors that are defined along polygonal lines in space (skew-polygonal)
and made of junctions of straight line segments. 
We designate by $U(y,x)$ a phase factor along 
the straight line segment $xy$, with an orientation from $x$ to $y$. 
A displacement of one end point of the segment, while the other one is 
fixed, generates a displacement of all points of the segment with 
appropriate weight factors. We may characterize this as representing a 
rigid path displacement. Parametrizing linearly the segment with a 
parameter $\lambda$, $0\le \lambda \le 1$, such that a point of the 
segment is represented as $z(\lambda)$, with $z(0)=x$ and $z(1)=y$, a 
displacement of one end point of the segment gives rise to two types of 
contribution, the first coming from the end point itself and the second 
coming from the inner points of the segment. One has for the rigid path 
derivatives the formulas
\be \lb{2e1}
\frac{\partial U(y,x)}{\partial y^{\alpha}}=-igA_{\alpha}(y)U(y,x)
+ig(y-x)^{\beta}\int_0^1d\lambda\,\lambda\, U(y,z(\lambda))F_{\beta\alpha}
(z(\lambda))U(z(\lambda),x),
\ee
\be \lb{2e2}
\frac{\partial U(y,x)}{\partial x^{\alpha}}=+igU(y,x)A_{\alpha}(x)
+ig(y-x)^{\beta}\int_0^1d\lambda\,(1-\lambda)\, U(y,z(\lambda))
F_{\beta\alpha}(z(\lambda))U(z(\lambda),x),
\ee
where $A$ is the gluon potential, $F$ its field strength, and $g$ the
coupling constant. In the above equations, the integrals represent the
inner contributions of the segment. When dealing with gauge invariant 
quantities, the end point contributions are usually cancelled by
similar contributions coming from neighboring segments or fields and 
it is the inner contributions of the segments that remain. We adopt 
for them the following notations: 
\bea 
\lb{2e3}
& &\frac{\bar\delta U(y,x)}{\bar\delta y^{\alpha +}}\equiv
ig(y-x)^{\beta}\int_0^1d\lambda\,\lambda\, U(y,z(\lambda))F_{\beta\alpha}
(z(\lambda))U(z(\lambda),x),\\
\lb{2e4}
& &\frac{\bar\delta U(y,x)}{\bar\delta x^{\alpha -}}\equiv
ig(y-x)^{\beta}\int_0^1d\lambda\,(1-\lambda)\, U(y,z(\lambda))F_{\beta\alpha}
(z(\lambda))U(z(\lambda),x).
\eea
Taking into account the orientation on $U$, the superscript $+$ or $-$ 
of the derivative variable indicates the segment on which it is acting
when we are in the presence of two joined segments. 
Thus if we have the expression $U(y,u)U(u,x)$, then the 
operator $\bar\delta/\bar\delta u^+$ will act on $U(u,x)$ only, through
the end point $u$ of the segment $xu$, while the operator
$\bar\delta/\bar\delta u^-$ will act on $U(y,u)$.
\par
The vacuum expectation value $W$ of a Wilson loop (or, equivalently, the 
Wilson loop average) along a closed polygonal contour with $n$ sides 
and $n$ junction points $x_1$, $x_2$, $\ldots$, $x_n$ will be denoted
$W_n$ and will be represented as the exponential of a functional  
$F_n$ \cite{mm,dv},  
\be \lb{2e5}
W_n=W(x_n,x_{n-1},\ldots,x_1)=
e^{{\displaystyle F_n(x_n,x_{n-1},\ldots,x_1)}}
=e^{{\displaystyle F_n}},
\ee
the orientation of the contour going from $x_1$ to $x_n$ through
$x_2$, $x_3$, etc. Then, the notation $\bar\delta F_n/\bar\delta x_i^- $ 
means that the derivative acts on the internal part of the segment 
$x_ix_{i+1}$ with $x_{i+1}$ held fixed ($x_{n+1}=x_1$), while 
$\bar\delta F_n/\bar\delta x_i^+$ means that the derivative acts on the 
internal part of the segment $x_{i-1}x_i$ with $x_{i-1}$ held fixed 
($x_0=x_n$).
\par
The 2PGIQGFs with phase factors along polygonal lines can be classified
according to the number of segments they contain. The 2PGIQGF with a
phase factor line with $n$ sides and $n-1$ junction points $t_1$, $t_2$, 
$\ldots$, $t_{n-1}$ between the segments is defined as
\be \lb{2e6}
S_{(n)}(x,x';t_{n-1},\dots,t_1)=-\frac{1}{N_c}\,\langle\overline \psi(x')
U(x',t_{n-1})U(t_{n-1},t_{n-2})\ldots U(t_1,x)\psi(x)\rangle,
\ee
the $\psi$s being the quark fields, with mass term $m$, belonging to the 
defining fundamental representation of the color gauge group $SU(N_c)$ 
and the vacuum averaging being defined in the path integral formalism. 
(Spinor indices are not written and the color indices are implicitly 
summed.) The orientation of the path in $S_{(n)}(x,x';t_{n-1},\dots,t_1)$ 
runs from $x$ to $x'$, passing by $t_1$, $t_2$, $\ldots$, $t_{n-1}$.
\par
The simplest 2PGIQGF corresponds to the case where $n=1$, for which the 
points $x$ and $x'$ are joined by a single straight line,
\be \lb{2e7}
S_{(1)}(x,x')\equiv S(x,x')=-\frac{1}{N_c}\,\langle\overline \psi(x')
\,U(x',x)\,\psi(x)\rangle.
\ee
(We shall generally omit the index 1 from that function.)
A graphical representation of the 2PGIQGFs $S_{(1)}$ and $S_{(3)}$ is 
shown in Fig. \rf{2f1}.
\par
\bfg
\bc
\begin{picture}(0,0)%
\includegraphics{2f1.pstex}%
\end{picture}%
\setlength{\unitlength}{3108sp}%
\begingroup\makeatletter\ifx\SetFigFont\undefined%
\gdef\SetFigFont#1#2#3#4#5{%
  \reset@font\fontsize{#1}{#2pt}%
  \fontfamily{#3}\fontseries{#4}\fontshape{#5}%
  \selectfont}%
\fi\endgroup%
\begin{picture}(6304,2073)(2101,-5440)
\put(2926,-5326){\makebox(0,0)[lb]{\smash{{\SetFigFont{9}{10.8}{\familydefault}{\mddefault}{\updefault}{\color[rgb]{0,0,0}$S_{(1)}$}%
}}}}
\put(2116,-5146){\makebox(0,0)[lb]{\smash{{\SetFigFont{9}{10.8}{\familydefault}{\mddefault}{\updefault}{\color[rgb]{0,0,0}$x$}%
}}}}
\put(4051,-5146){\makebox(0,0)[lb]{\smash{{\SetFigFont{9}{10.8}{\familydefault}{\mddefault}{\updefault}{\color[rgb]{0,0,0}$x'$}%
}}}}
\put(5716,-5146){\makebox(0,0)[lb]{\smash{{\SetFigFont{9}{10.8}{\familydefault}{\mddefault}{\updefault}{\color[rgb]{0,0,0}$x$}%
}}}}
\put(8236,-5146){\makebox(0,0)[lb]{\smash{{\SetFigFont{9}{10.8}{\familydefault}{\mddefault}{\updefault}{\color[rgb]{0,0,0}$x'$}%
}}}}
\put(6076,-3796){\makebox(0,0)[lb]{\smash{{\SetFigFont{9}{10.8}{\familydefault}{\mddefault}{\updefault}{\color[rgb]{0,0,0}$t_1$}%
}}}}
\put(7831,-3526){\makebox(0,0)[lb]{\smash{{\SetFigFont{9}{10.8}{\familydefault}{\mddefault}{\updefault}{\color[rgb]{0,0,0}$t_2$}%
}}}}
\put(6841,-5371){\makebox(0,0)[lb]{\smash{{\SetFigFont{9}{10.8}{\familydefault}{\mddefault}{\updefault}{\color[rgb]{0,0,0}$S_{(3)}$}%
}}}}
\end{picture}%

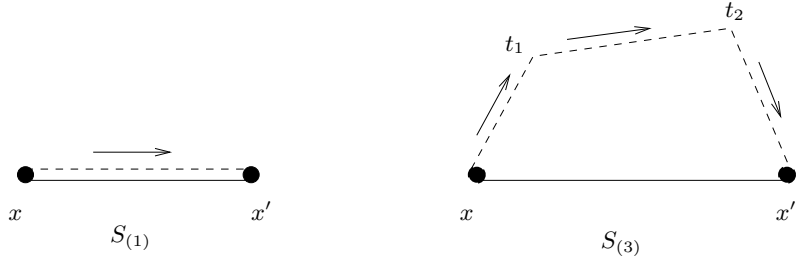
\caption{Graphical representation of the 2PGIQGFs $S_{(1)}$ and $S_{(3)}$.
The solid lines represent the quark field contractions, the dotted
lines the phase factor along the polygonal lines and the arrows the 
orientation on them.}
\lb{2f1}
\ec
\efg
For the internal parts of rigid path derivatives, we have the
definitions
\be \lb{2e8}
\frac{\bar\delta S_{(n)}(x,x';t_{n-1},\dots,t_1)}{\bar\delta x^{\mu -}}
=-\frac{1}{N_c}\,\langle\overline \psi(x')\,
U(x',t_{n-1})U(t_{n-1},t_{n-2})\ldots 
\frac{\bar\delta U(t_1,x)}{\bar\delta x^{\mu -}}\,\psi(x)\rangle,
\ee
\be \lb{2e9}
\frac{\bar\delta S_{(n)}(x,x';t_{n-1},\dots,t_1)}{\bar\delta x^{'\nu +}}
=-\frac{1}{N_c}\,\langle\overline \psi(x')\,
\frac{\bar\delta U(x',t_{n-1})}{\bar\delta x^{'\nu +}}
U(t_{n-1},t_{n-2})\ldots U(t_1,x)\,\psi(x)\rangle.
\ee
\par
$S_{(1)}$ and $S_{(n)}$ satisfy the following equations of motion:
\bea 
\lb{2e10}
& &(i\gamma.\partial_{(x)}-m)S(x,x')=i\delta^4(x-x')
+i\gamma^{\mu}\,\frac{\bar\delta S(x,x')}{\bar\delta x^{\mu -}},\\
\lb
{2e11}
& &S(x,x')(-i\gamma.\stackrel{\leftarrow}{\partial}_{(x')}-m)=
i\delta^4(x-x')
-i\frac{\bar\delta S(x,x')}{\bar\delta x^{'\mu +}}\,\gamma^{\mu},
\eea
\bea \lb{2e12}
& &(i\gamma.\partial_{(x)}-m)S_{(n)}(x,x';t_{n-1},\ldots,t_1)=
i\delta^4(x-x')e^{{\displaystyle F_{n}(x,t_{n-1},\ldots,t_1)}}
\nonumber \\
& &\ \ \ \ \ \ \ \ \ \
+i\gamma^{\mu}\frac{\bar\delta S_{(n)}(x,x';t_{n-1},\ldots,t_1)}
{\bar\delta x^{\mu -}},\\
\lb{2e13}
& &S_{(n)}(x,x';t_{n-1},\ldots,t_1)
(-i\gamma.\stackrel{\leftarrow}{\partial}_{(x')}-m)=
i\delta^4(x-x')e^{{\displaystyle F_{n}(x,t_{n-1},\ldots,t_1)}}
\nonumber \\
& &\ \ \ \ \ \ \ \ \ \
-i\frac{\bar\delta S_{(n)}(x,x';t_{n-1},\ldots,t_1)}
{\bar\delta x^{'\mu +}}\,\gamma^{\mu}.
\eea
\par
The $S_{(n)}$s also satisfy equations related to the junction
points $t_1$, $\ldots$, $t_{n-1}$ on the polygonal line or more
generally to local deformations of the paths. These involve the 
gluon field equations of motion and lead to equations
related to the properties of phase factors and Wilson loops
\cite{p,mm,mgd,mk}. They should mainly be used for the determination
of the expressions of the Wilson loop averages. In the present paper,
the latter are assumed to be known and therefore the corresponding 
equations will not be considered.
\par   
Multiplying the equations of motion (\rf{2e12}) and (\rf{2e13}) with 
$S(t_1,x)$ and $S(x',t_{n-1})$, respectively, and integrating with 
respect to $x$ or $x'$, one can establish functional relations between 
the various 2PGIQGFs. For $S_{(n)}$, one has
\bea \lb{2e14}
& &S_{(n)}(x,x';t_{n-1},\ldots,t_1)=
S(x,x')\,e^{{\displaystyle F_{n+1}(x',t_{n-1},\ldots,t_1,x)}}
\nonumber \\
& &\ \ \ \ +\Big(\frac{\bar\delta S(x,y_1)}{\bar\delta y_1^{\alpha_1 +}}
+S(x,y_1)\frac{\bar\delta }{\bar\delta y_1^{\alpha_1 -}}\Big)
\,\gamma^{\alpha_1}\,S_{(n+1)}(y_1,x';t_{n-1},\ldots,t_1,x)\nonumber \\
& &\ \ \ \ \ \ \ \ \ \
=S(x,x')\,e^{{\displaystyle F_{n+1}(x',t_{n-1},\ldots,t_1,x)}}
\nonumber \\
& &\ \ \ \ -S_{(n+1)}(x,z_1;x',t_{n-1},\ldots,t_1)\,\gamma^{\beta_1}\,
\Big(\frac{\bar\delta S(z_1,x')}{\bar\delta z_1^{\beta_1 -}}
+\frac{\stackrel{\leftarrow}{\bar\delta}}
{\bar\delta z_1^{\beta_1 +}}S(z_1,x')\Big).
\eea
(Integrations on intermediate variables are implicit and will not be
written throughout this paper. Here, $y_1$ and $z_1$ are integration
variables.) By iterating these equations with respect to the higher 
$S_{(n)}$s of the right-hand sides and assuming that the last term 
rejected to infinity tends to zero, one ends up with a series expansion 
of any $S_{(n)}$ ($n>1$) in terms of $S$ and its derivative and 
derivatives of logarithms of Wilson loop averages. This shows that among 
the infinite set of 2PGIQGFs with polygonal lines, only the first one, with 
one single straight line, is a genuine dynamical independent quantity;
all others are in principle calculable from it, provided one knows to 
evaluate the rigid path derivative of $S$ and the Wilson loop averages.
\par
The calculation of $S$ proceeds from the equation of motion 
(\rf{2e10}) [or (\rf{2e11})]. It is then necessary to devise a method
for evaluating the rigid path derivative 
$\bar\delta S(x,x')/\bar\delta x^{\mu -}$. This is done by applying
the rigid path derivative operator on both sides of Eq. (\rf{2e14})
and repeating iteration operations. Specializing the result for $S$,
one finds
\bea \lb{2e15}
& &\frac{\bar\delta S(x,x')}{\bar\delta x^{\mu -}}=
\frac{\bar\delta F_{2}(x',x)}{\bar\delta x^{\mu -}}\,S(x,x')
-\frac{\bar\delta^2 F_3(x',x,y_1)}{\bar\delta x^{\mu -}
\bar\delta y_1^{\alpha_1 +}}\,S(x,y_1)\,\gamma^{\alpha_1}\,
S_{(2)}(y_1,x';x)
\nonumber \\
& &\  -\sum_{n=3}^{\infty}\Big(\frac{\bar\delta S(x,y_1)}
{\bar\delta y_1^{\alpha_1 +}}+S(x,y_1)\frac{\bar\delta}
{\bar\delta y_1^{\alpha_1 -}}\Big)\,\gamma^{\alpha_1}\nonumber \\
& &\ \   
\times\cdots\times\,\Big(\frac{\bar\delta S(y_{n-3},y_{n-2})}
{\bar\delta y_{n-2}^{\alpha_{n-2} +}}+S(y_{n-3},y_{n-2})
\frac{\bar\delta}{\bar\delta y_{n-2}^{\alpha_{n-2} -}}\Big)
\gamma^{\alpha_{n-2}}\,\nonumber \\
& &\ \ \times\, 
\frac{\bar\delta^2 F_{n+1}(x',x,y_1,\ldots,y_{n-1})}{\bar\delta x^{\mu -}
\bar\delta y_{n-1}^{\alpha_{n-1} +}}\,S(y_{n-2},y_{n-1})\,
\gamma^{\alpha_{n-1}}\,S_{(n)}(y_{n-1},x';x,y_1,\ldots,y_{n-2}).
\nonumber \\
& &
\eea
\par
The right-hand side involves a series of terms in which the $n$th-order
one contains $S_{(n)}$ and a Wilson loop with a polygonal contour with
$(n+1)$ sides. One notices from the locations of the second-order 
derivatives acting on the $F$s the absence of reducible-type contributions 
in the corresponding expressions; the latter are expected to be 
part of the definition of the $S_{(n)}$s when expressed in terms of free 
propagators. The calculation should be completed by 
bringing all derivative operators to the right; the final form shows
that the $n$th-order term contains globally $n$ derivatives acting on
the logarithm of the corresponding Wilson loop average and/or on the Green's
function $S$ (at most at first order for the latter). Furthermore, each 
derivative acting on the Wilson loop operates on a different segment from 
the others; this prevents the appearance of singularities arising from 
derivatives acting on the same point. The Wilson loop contributions have 
the characteristics of being irreducible and are classified into the
following three categories: connected, crossed and nested \cite{s1}. 
The form (\rf{2e15}) is the most convenient one for comparisons with other 
cases, such as those of the 4PGIQGFs. 
\par
Equation (\rf{2e15}) can be considered as the analogue of the self-energy 
Dyson-Schwinger equation in the case of the ordinary Green's function 
\cite{d,schw}. Defining the latter as 
$\widetilde{S}(x,x')=\frac{1}{N_c}\,\langle \psi(x)\,
\overline\psi(x')\rangle$, its equation of motion takes the form
\be \lb{2e16}
(i\gamma.\partial_{(x)}-m)\widetilde{S}(x,x')=i\delta^4(x-x')
+\Sigma(x,y)\,\widetilde{S}(y,x'),
\ee
where $\Sigma$ defines the self-energy; it is a functional of the 
Green's function $\widetilde S$ itself, together with other Green's
functions of interest, like that of the gluon field or of the photon 
field in the case of QED. The expression of $\Sigma[\widetilde{S}]$
in terms of $\widetilde{S}$ and the other two-point Green's functions
defines the Dyson-Schwinger equation.
\par  
In the present case, however, one meets a more complicated situation 
in two respects. First, Eq. (\rf{2e15}) involves in its right-hand side
the whole set of 2PGIQGFs defined along polygonal lines, although all of 
them are ultimately expressible in terms of $S$. This feature, on the
other hand, is an indication that the set of 2PGIQGFs along polygonal lines
is closed, since no other types of contour are needed to reach the
final equation for $S$. Second, the integrals that are present are not
of the convolution type; they overlap all terms that
accompany them, which is due to the presence of the Wilson loops, whose
contours pass by all points that are present. Taking into account these
facts, one needs to introduce matrix-type self-energy operators, 
$\Sigma_{mn}$, where the first index refers to the initial Green's
function that is considered ($S_{(m)}$) and the second one to the Green's
function $S_{(n)}$ that appears in the right-hand side of the equation.
With this definition, the equation of motion (\rf{2e10}) can be
schematically written in the form
\be \lb{2e17}
(i\gamma.\partial_{(x)}-m)S(x,x')=i\delta^4(x-x')
+\sum_{n=1}^{\infty}\,\Big(\Sigma_{1n}[S]*S_{(n)}\Big),
\ee
where the star operation represents the integrals involved in the
term containing $S_{(n)}$ in the right-hand side of Eq. (\rf{2e15}),
and the functional expression of $\Sigma_{1n}[S]$ is deduced from
that equation by identification. 
\par
The above results can also be applied with obvious transpositions to
the equation of motion (\rf{2e11}). The evaluation of 
$\bar\delta S(x,x')/\bar\delta x^{'\mu +}$ can be done in two 
ways. First, one might use the second expression of the functional
relationship of Eq. (\rf{2e14}). In this case the iterative expansion
is done in the reverse order to that of Eq. (\rf{2e15}). The corresponding
equation takes then the schematic form
\be \lb{2e18}
S(x,x')(-i\gamma.\stackrel{\leftarrow}{\partial}_{(x')}-m)=
i\delta^4(x-x')
+\sum_{n=1}^{\infty}\,\Big(S_{(n)}*\widetilde{\Sigma}_{1n}[S]\Big),
\ee
where $\widetilde{\Sigma}_{1n}[S]$ is deduced from 
${\Sigma}_{1n}[S]$ by reversing the orders of appearance of various
matrices and Green's functions and some of their arguments, changing the
sign of terms containing an odd number of explicit $\gamma$ matrices and 
replacing the rigid path derivatives 
$\bar\delta/\bar\delta x^{\mu -}$ and 
$\bar\delta/\bar\delta y_j^{\alpha_j +}$ with
$\bar\delta/\bar\delta x^{'\mu +}$ and
$\bar\delta/\bar\delta z_j^{\beta_j -}$, respectively.
Second, one might use the same functional relations as for obtaining
Eq. (\rf{2e15}). In this case, the only modification is the 
replacement of the the rigid path derivative 
$\bar\delta/\bar\delta x^{\mu -}$ by 
$\bar\delta/\bar\delta x^{'\mu +}$, while in Eq. (\rf{2e11})
we have to take into account the global change of sign in front of
$\bar\delta S(x,x')/\bar\delta x^{'\mu +}$ and the position
of the matrix $\gamma^{\mu}$ on the utmost right. We shall
write the resulting expression of the equation in the form
\be \lb{2e19}
S(x,x')(-i\gamma.\stackrel{\leftarrow}{\partial}_{(x')}-m)=
i\delta^4(x-x')
+\sum_{n=1}^{\infty}\,\Big(\widehat{\Sigma}_{1n}[S]*S_{(n)}\Big).
\ee
\par
It is to be emphasized that once the operator $\Sigma_{1n}[S]$ of
Eq. (\rf{2e17}) [or its analogue of Eqs. (\rf{2e18}) and (\rf{2e19})]
has been evaluated by means of Eq. (\rf{2e15}), then the Green's
functions $S_{(1)}$, $\ldots$, $S_{(n)}$, etc., have to be
considered as ordinary complex functions of the variables $x$, $x'$,
$y_1$, $\ldots$, $y_n$, etc., satisfying translation invariance 
and well defined Lorentz transformation properties. The information
contained in their phase factors along the rigid straight line 
segments and polygonal lines is now expressed by means of the 
corresponding Wilson loop averages and their rigid path derivatives, 
which, after evaluation, are themselves ordinary functions of their 
arguments $x$, $x'$, etc. (the junction points of the segments).  
The straight line segments and the polygonal lines no longer 
introduce additional degrees of freedom, since their geometry is 
completely determined by the knowledge of the positions of the 
junction points of the segments. In particular, the operator 
$\partial/\partial x$ of the Dirac operator in Eq. (\rf{2e17})
acts as an ordinary derivative operator on $S$. This is why Eq. 
(\rf{2e17}) and its analogues have the status of integro\-differential
equations.
\par 
In the rest of the paper we shall consider systems involving two quarks 
with different flavors and generally with different masses. To distinguish
their individual Green's functions we shall introduce an additional index
for their notation. Thus, they will be denoted $S_{1,(n)}$ and 
$S_{2,(n)}$, respectively, corresponding to quark 1 and quark 2. 
The simplest Green's functions (with one straight line) $S$
($\equiv S_{(1)}$) will be denoted $S_1$ and $S_2$.
\par  

\section{Four-point Green's functions} \lb{s3}
\setcounter{equation}{0}

We now consider two different quark fields, labeled with indices 1 
and 2, respectively, with mass terms $m_1$ and $m_2$. Four-point GIQGFs 
are constructed by including gluon field phase factors between the quark 
and the antiquark fields. Considering a polygonal line made of $n$
segments and another made of one segment, we define the 4PGIQGF $G_{(n)}$
as
\bea \lb{3e1}
& &G_{(n)\alpha\beta,\beta'\alpha'}
(x_1,x_2;x_2',x_1';t_{n-1},t_{n-2},\ldots,t_1)
\nonumber \\
& & 
=-\frac{1}{N_c}\,\langle\overline \psi_{2\beta}(x_2)
U(x_2,t_{n-1})\ldots U(t_1,x_1)\psi_{1\alpha}(x_1)
\overline \psi_{1\alpha'}(x_1')U(x_1',x_2')\psi_{2\beta'}(x_2')\rangle,
\eea
where $\alpha,\beta,\beta',\alpha'$ are the spinor indices of the 
quark fields.
\par
The simplest such Green's function is $G_{(1)}$,
\be \lb{3e2}
G_{(1)}(x_1,x_2;x_2',x_1')=
-\frac{1}{N_c}\,\langle\overline \psi_2(x_2)
U(x_2,x_1)\psi_1(x_1)
\overline \psi_1(x_1')U(x_1',x_2')\psi_2(x_2')\rangle.
\ee
\par
A graphical representation of $G_{(1)}$ and $G_{(3)}$ is shown
in Fig. \rf{3f1}.
\par
\bfg
\bc
\begin{picture}(0,0)%
\includegraphics{3f1.pstex}%
\end{picture}%
\setlength{\unitlength}{2693sp}%
\begingroup\makeatletter\ifx\SetFigFont\undefined%
\gdef\SetFigFont#1#2#3#4#5{%
  \reset@font\fontsize{#1}{#2pt}%
  \fontfamily{#3}\fontseries{#4}\fontshape{#5}%
  \selectfont}%
\fi\endgroup%
\begin{picture}(9698,3423)(1381,-5845)
\put(8776,-5776){\makebox(0,0)[lb]{\smash{{\SetFigFont{8}{9.6}{\familydefault}{\mddefault}{\updefault}{\color[rgb]{0,0,0}$G_{(3)}$}%
}}}}
\put(2701,-5776){\makebox(0,0)[lb]{\smash{{\SetFigFont{8}{9.6}{\familydefault}{\mddefault}{\updefault}{\color[rgb]{0,0,0}$G_{(1)}$}%
}}}}
\put(4456,-2626){\makebox(0,0)[lb]{\smash{{\SetFigFont{8}{9.6}{\familydefault}{\mddefault}{\updefault}{\color[rgb]{0,0,0}$x_1'$}%
}}}}
\put(1396,-4921){\makebox(0,0)[lb]{\smash{{\SetFigFont{8}{9.6}{\familydefault}{\mddefault}{\updefault}{\color[rgb]{0,0,0}$x_2$}%
}}}}
\put(7246,-5326){\makebox(0,0)[lb]{\smash{{\SetFigFont{8}{9.6}{\familydefault}{\mddefault}{\updefault}{\color[rgb]{0,0,0}$x_2$}%
}}}}
\put(10486,-5101){\makebox(0,0)[lb]{\smash{{\SetFigFont{8}{9.6}{\familydefault}{\mddefault}{\updefault}{\color[rgb]{0,0,0}$x_2'$}%
}}}}
\put(6886,-3346){\makebox(0,0)[lb]{\smash{{\SetFigFont{8}{9.6}{\familydefault}{\mddefault}{\updefault}{\color[rgb]{0,0,0}$t_1$}%
}}}}
\put(1576,-2851){\makebox(0,0)[lb]{\smash{{\SetFigFont{8}{9.6}{\familydefault}{\mddefault}{\updefault}{\color[rgb]{0,0,0}$x_1$}%
}}}}
\put(4771,-5146){\makebox(0,0)[lb]{\smash{{\SetFigFont{8}{9.6}{\familydefault}{\mddefault}{\updefault}{\color[rgb]{0,0,0}$x_2'$}%
}}}}
\put(10981,-2851){\makebox(0,0)[lb]{\smash{{\SetFigFont{8}{9.6}{\familydefault}{\mddefault}{\updefault}{\color[rgb]{0,0,0}$x_1'$}%
}}}}
\put(7696,-2581){\makebox(0,0)[lb]{\smash{{\SetFigFont{8}{9.6}{\familydefault}{\mddefault}{\updefault}{\color[rgb]{0,0,0}$x_1$}%
}}}}
\put(6616,-4381){\makebox(0,0)[lb]{\smash{{\SetFigFont{8}{9.6}{\familydefault}{\mddefault}{\updefault}{\color[rgb]{0,0,0}$t_2$}%
}}}}
\end{picture}%

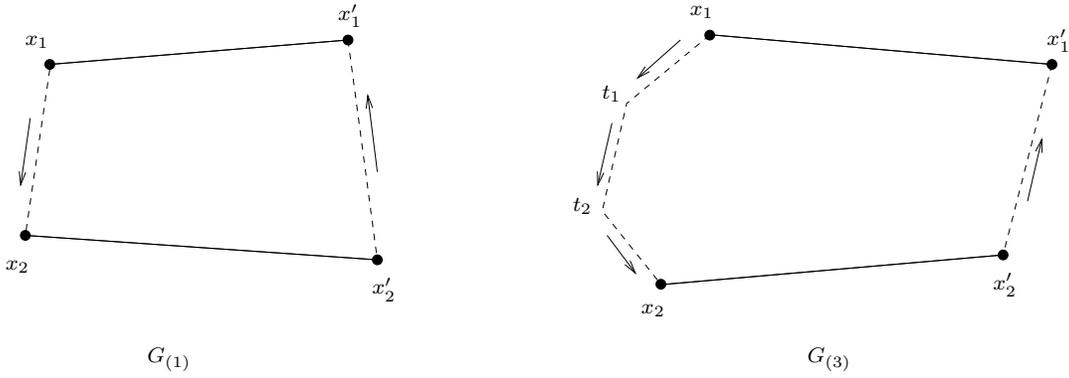
\caption{Graphical representation of the 4PGIQGFs $G_{(1)}$ and $G_{(3)}$.
Same conventions as in Fig. \rf{2f1}.}
\lb{3f1}
\ec
\efg
In the following, for the study of the bound state problem, 
the points $x_1'$ and $x_2'$ will be sent to $-\infty$ in time;
they will then disappear by factorization from the bound
state equations, which is why the classification of the 4PGIQGFs 
hinges here only on the line between the points $x_1$ and $x_2$.
\par
The 4PGIQGFs satisfy the following equations of motion:
\bea 
\lb{3e3}
& &(i\gamma.\partial_1-m_1)G_{(n)}(x_1,x_2;x_2',x_1';t_{n-1},\ldots,t_1)
\nonumber \\
& &\ \ \ \ \ \ \
=i\delta^4(x_1-x_1')S_{2,(n+1)}(x_2',x_2;t_{n-1},\ldots,t_1,x_1)
\nonumber \\
& &\ \ \ \ \ \ \ \ \ \
+i\gamma^{\mu}\frac{\bar\delta}{\bar\delta x_1^{\mu -}} 
G_{(n)}(x_1,x_2;x_2',x_1';t_{n-1},\ldots,t_1)\Big|_{x_1t_1}^{},
\eea
\bea
\lb{3e4}
& &G_{(n)}(x_1,x_2;x_2',x_1';t_{n-1},\ldots,t_1)
(-i\gamma.\stackrel{\leftarrow}{\partial}_2-m_2)
\nonumber \\
& &\ \ \ \ \ \ \
=i\delta^4(x_2-x_2')S_{1,(n+1)}(x_1,x_1';x_2,t_{n-1},\ldots,t_1)
\nonumber \\
& &\ \ \ \ \ \ \ \ \ \
-i\frac{\bar\delta}{\bar\delta x_2^{\nu +}} 
G_{(n)}(x_1,x_2;x_2',x_1';t_{n-1},\ldots,t_1)
\Big|_{t_{n-1}x_2}^{}\,\gamma^{\nu}.
\eea
($\partial_1=\partial/\partial x_1$ and 
$\partial_2=\partial/\partial x_2$.) The $\gamma$ matrices that act on
$G$ from the left, act on its first spinor index ($\alpha$), while 
those acting from the right act on its second spinor index ($\beta$),
as defined in Eq. (\rf{3e1}). 
\par
Multiplying Eqs. (\rf{3e3}) and (\rf{3e4}) with $S_1(t_1,x_1)$ and 
$S_2(x_2,t_{n-1})$, respectively, and integrating, one obtains the
following functional relations between different 4PGIQGFs:
\bea \lb{3e5}
& &G_{(n)}(x_1,x_2;x_2',x_1';t_{n-1},\ldots,t_1)=
S_1(x_1,x_1')S_{2,(n+2)}(x_2',x_2;t_{n-1},\ldots,t_1,x_1,x_1')
\nonumber \\
& &\ \ \ \ +\Big(\frac{\bar\delta S_1(x_1,y_1)}
{\bar\delta y_1^{\alpha_1 +}}
+S_1(x_1,y_1)\frac{\bar\delta }{\bar\delta y_1^{\alpha_1 -}}\Big)
\,\gamma^{\alpha_1}\,G_{(n+1)}(y_1,x_2;x_2',x_1';t_{n-1},\ldots,t_1,x_1)
\nonumber \\
& &\ \ \ \ \ \ \ \ \ \
=S_{1,(n+2)}(x_1,x_1';x_2',x_2,t_{n-1},\ldots,t_1)S_2(x_2',x_2)
\nonumber \\
& &\ \ \ \ -G_{(n+1)}(x_1,z_1;x_2',x_1';x_2,t_{n-1},\ldots,t_1)
\,\gamma^{\beta_1}\,
\Big(\frac{\bar\delta S_2(z_1,x_2)}{\bar\delta z_1^{\beta_1 -}}
+\frac{\stackrel{\leftarrow}{\bar\delta}}
{\bar\delta z_1^{\beta_1 +}}S_2(z_1,x_2)\Big).\nonumber \\
& &
\eea
\par

\section{Bound state equations} \lb{s4}
\setcounter{equation}{0}

In order to obtain bound state equations, one considers in the
4PGIQGFs the limit of large timelike separations between the set of 
points ($x_1,t_1,\ldots,t_{n-1},x_2$) and the set ($x_2',x_1'$). 
In this limit, the Green's functions can be saturated by a complete
set of hadronic states, among which are single mesons representing
bound states of quarks and antiquarks. To simplify the analysis,
one may either consider the large-$N_c$ limit, in which case only
single poles survive \cite{th1,wt}, or simply neglect inelasticity
effects to ensure the stability of the bound state. By appropriate
projection operations \cite{gml} or limiting procedures to the pole
position in the total momentum space, it is possible to select one
particular bound state among the whole set of intermediate states.
\par
We shall refer to the selected bound state with its total four-momentum 
$P$ only, discarding other quantum numbers, which will not play any role 
in the following. (We assume that the bound state is nondegenerate.)   
The wave functions are classified according to the number of straight
line segments existing on the polygonal lines of the phase factors
($n,j=1,2,\ldots,\infty$),
\be \lb{4e1}
\Phi_{(n)\alpha\beta}(x_1,x_2;t_{n-1},\ldots,t_1)=
-\frac{1}{\sqrt{N_c}}<0|T(\overline \psi_{2\beta}(x_2)
U(x_2,t_{n-1})\ldots U(t_1,x_1)\psi_{1\alpha}(x_1))|P>,
\ee
\be \lb{4e2}
\overline\Phi_{(j)\beta'\alpha'}(x_2',x_1';t_{j-1},\ldots,t_1)=
\frac{1}{\sqrt{N_c}}<P|T(\overline \psi_{1\alpha'}(x_1')
U(x_1',t_{j-1})\ldots U(t_1,x_2')\psi_{2\beta'}(x_2'))|0>.
\ee
They become in the simplest cases $n=1$ and $j=1$,
\bea
\lb{4e3}
& &\Phi_{(1)}(x_1,x_2)=
-\frac{1}{\sqrt{N_c}}<0|T(\overline \psi_2(x_2)
U(x_2,x_1)\psi_1(x_1))|P>,\\
\lb{4e4}
& &\overline\Phi_{(1)}(x_2',x_1')=
\frac{1}{\sqrt{N_c}}<P|T(\overline \psi_1(x_1')
U(x_1',x_2')\psi_2(x_2'))|0>.
\eea
(The dependence of the wave functions on the total four-momentum $P$ 
of the bound state is omitted from their arguments for notational 
simplification.) We note that the above wave functions 
for all $n$s describe the same bound state, but differ in their 
expressions due to differences in their contents with respect 
to the phase factor lines.   
\par
Taking large timelike separations between initial and final 
coordinates as described above, or going in total momentum space
to the pole position, and selecting one bound state,
the equations of motion (\rf{3e3}) and (\rf{3e4}) of the 4PGIQGFs
are transformed into wave equations (their inhomogeneous parts
not contributing to the bound states poles),
\be \lb{4e5}
(i\gamma.\partial_1-m_1)\Phi_{(n)}(x_1,x_2;t_{n-1},\ldots,t_1)
=+i\gamma^{\mu}\frac{\bar\delta}{\bar\delta x_1^{\mu -}} 
\Phi_{(n)}(x_1,x_2;t_{n-1},\ldots,t_1)\Big|_{x_1t_1}^{},\\
\ee
\be \lb{4e6}
\Phi_{(n)}(x_1,x_2;t_{n-1},\ldots,t_1)
(-i\gamma.\stackrel{\leftarrow}{\partial}_2-m_2)
=-i\frac{\bar\delta}{\bar\delta x_2^{\nu +}} 
\Phi_{(n)}(x_1,x_2;t_{n-1},\ldots,t_1)
\Big|_{t_{n-1}x_2}^{}\,\gamma^{\nu},
\ee
which become for the case $n=1$
\bea 
\lb{4e7}
& &(i\gamma.\partial_1-m_1)\Phi_{(1)}(x_1,x_2)
=+i\gamma^{\mu}\frac{\bar\delta}{\bar\delta x_1^{\mu -}} 
\Phi_{(1)}(x_1,x_2),\\
\lb{4e8}
& &\Phi_{(1)}(x_1,x_2)
(-i\gamma.\stackrel{\leftarrow}{\partial}_2-m_2)
=-i\frac{\bar\delta}{\bar\delta x_2^{\nu +}} 
\Phi_{(1)}(x_1,x_2)\,\gamma^{\nu}.
\eea
\par
A wave function $\Phi_{(n)}$ thus satisfies two independent
Dirac-type equations. They should, however, be compatible among 
themselves in order not to give rise to a vanishing solution. 
The compatibility condition is obtained by making the two Dirac 
operators act on the wave function in different orders and subtracting 
the corresponding results from each other. Since the two Dirac operators 
commute among themselves, the result should be zero. One finds
\be \lb{4e9}
\Big(\frac{\bar\delta}{\bar\delta x_2^{\nu +}}
\frac{\bar\delta}{\bar\delta x_1^{\mu -}}
-\frac{\bar\delta}{\bar\delta x_1^{\mu -}}
\frac{\bar\delta}{\bar\delta x_2^{\nu +}}\Big)
\Phi_{(n)}=0.
\ee
For $n\ge 2$, the commutativity of the two rigid path derivatives
results from the fact that they operate on different segments in 
an uncorrelated way. For $n=1$, since they operate on 
the same segment, they may act on the same point, giving rise to
additional singularities. It can, however, be shown that because
of the Bianchi identities satisfied by the gluon fields, even in this 
case the two operators commute \cite{js}. The two wave equations
(\rf{4e5}) and (\rf{4e6}), or (\rf{4e7}) and (\rf{4e8}), are therefore
compatible among themselves. 
\par
In order to evaluate the interaction part of the wave equations,
it is necessary to express the rigid path derivatives in terms of
calculable kernels. We follow here a method similar to that used for 
the 2PGIQGFs (Sec. \rf{s2}). By selecting in Eqs. (\rf{3e5}) the
bound state contribution, one obtains the following two equivalent 
equations, expressing a wave function $\Phi_{(n)}$ in terms of 
$\Phi_{(n+1)}$ and the 2PGIQGFs $S_1$ or $S_2$:
\bea \lb{4e10}
& &\Phi_{(n)}(x_1,x_2;t_{n-1},\ldots,t_1)=
\nonumber \\
& &\ \ \ +\Big(\frac{\bar\delta S_1(x_1,y_1)}
{\bar\delta y_1^{\alpha_1 +}}
+S_1(x_1,y_1)\frac{\bar\delta }{\bar\delta y_1^{\alpha_1 -}}\Big)
\,\gamma^{\alpha_1}\,\Phi_{(n+1)}(y_1,x_2;t_{n-1},\ldots,t_1,x_1)
\nonumber \\
& &\ \ \ 
=-\Phi_{(n+1)}(x_1,z_1;x_2,t_{n-1},\ldots,t_1)
\,\gamma^{\beta_1}\,
\Big(\frac{\bar\delta S_2(z_1,x_2)}{\bar\delta z_1^{\beta_1 -}}
+\frac{\stackrel{\leftarrow}{\bar\delta}}
{\bar\delta z_1^{\beta_1 +}}S_2(z_1,x_2)\Big).
\eea
However, contrary to the 2PGIQGF case [Eqs. (\rf{2e14})], they 
do not contain the lowest-index wave function $\Phi_{(1)}$,
which could generate an iterative series and allow for the 
calculation of the $\Phi_{(n)}$s in terms of $\Phi_{(1)}$. The
difficulty can be overcome by adding to the above equations identities 
involving $\Phi_{(1)}$. Considering the equations of motion 
(\rf{2e12}) and (\rf{2e13}), multiplying them with $\Phi_{(1)}(t_1,x_1)$ 
and $\Phi_{(1)}(x_2,t_{n-1})$, respectively, and integrating, one 
obtains the two equations,
\newpage
\bea 
\lb{4e11}
& &\Phi_{(1)}(x_1,x_2)\,
e^{{\displaystyle F_{n+1}(x_2,t_{n-1},\ldots,t_1,x_1)}}
\nonumber \\
& &\ \ \ \ 
+\Big(\frac{\bar\delta \Phi_{(1)}(x_1,y_1)}{\bar\delta y_1^{\alpha_1 +}}
+\Phi_{(1)}(x_1,y_1)\frac{\bar\delta }{\bar\delta y_1^{\alpha_1 -}}\Big)
\,\gamma^{\alpha_1}\,S_{2,(n+1)}(y_1,x_2;t_{n-1},\ldots,t_1,x_1)=0,
\nonumber \\
& & \\ 
\lb{4e12}
& &\Phi_{(1)}(x_1,x_2)\,
e^{{\displaystyle F_{n+1}(x_2,t_{n-1},\ldots,t_1,x_1)}}
\nonumber \\
& &\ \ \ \
-S_{1,(n+1)}(x_1,z_1;x_2,t_{n-1},\ldots,t_1)\,\gamma^{\beta_1}\,
\Big(\frac{\bar\delta \Phi_{(1)}(z_1,x_2)}{\bar\delta z_1^{\beta_1 -}}
+\frac{\stackrel{\leftarrow}{\bar\delta}}
{\bar\delta z_1^{\beta_1 +}}\Phi_{(1)}(z_1,x_2)\Big)=0.\nonumber \\
& &
\eea
These can now be added, respectively, to the two expressions of 
$\Phi_{(n)}$ in Eqs. (\rf{4e10}), yielding the following new
functional relations:
\bea 
\lb{4e13}
& &\Phi_{(n)}(x_1,x_2;t_{n-1},\ldots,t_1)=
\Phi_{(1)}(x_1,x_2)\,
e^{{\displaystyle F_{n+1}(x_2,t_{n-1},\ldots,t_1,x_1)}}
\nonumber \\
& &\ \ \ \
+\Big(\frac{\bar\delta \Phi_{(1)}(x_1,y_1)}{\bar\delta y_1^{\alpha_1 +}}
+\Phi_{(1)}(x_1,y_1)\frac{\bar\delta }{\bar\delta y_1^{\alpha_1 -}}\Big)
\,\gamma^{\alpha_1}\,S_{2,(n+1)}(y_1,x_2;t_{n-1},\ldots,t_1,x_1) 
\nonumber \\
& &\ \ \ \ +\Big(\frac{\bar\delta S_1(x_1,y_1)}
{\bar\delta y_1^{\alpha_1 +}}
+S_1(x_1,y_1)\frac{\bar\delta }{\bar\delta y_1^{\alpha_1 -}}\Big)
\,\gamma^{\alpha_1}\,\Phi_{(n+1)}(y_1,x_2;t_{n-1},\ldots,t_1,x_1),
\eea
\bea
\lb{4e14}
& &\Phi_{(n)}(x_1,x_2;t_{n-1},\ldots,t_1)=
\Phi_{(1)}(x_1,x_2)\,
e^{{\displaystyle F_{n+1}(x_2,t_{n-1},\ldots,t_1,x_1)}}
\nonumber \\
& &\ \ \ \
-S_{1,(n+1)}(x_1,z_1;x_2,t_{n-1},\ldots,t_1)\,\gamma^{\beta_1}\,
\Big(\frac{\bar\delta \Phi_{(1)}(z_1,x_2)}{\bar\delta z_1^{\beta_1 -}}
+\frac{\stackrel{\leftarrow}{\bar\delta}}
{\bar\delta z_1^{\beta_1 +}}\Phi_{(1)}(z_1,x_2)\Big)
\nonumber \\
& &\ \ \ \
-\Phi_{(n+1)}(x_1,z_1;x_2,t_{n-1},\ldots,t_1)\,\gamma^{\beta_1}\,
\Big(\frac{\bar\delta S_2(z_1,x_2)}{\bar\delta z_1^{\beta_1 -}}
+\frac{\stackrel{\leftarrow}{\bar\delta}}
{\bar\delta z_1^{\beta_1 +}}S_2(z_1,x_2)\Big).
\eea
They can be used to express, through iterative calculations, 
$\Phi_{(n)}$ in terms of $\Phi_{(1)}$, $S_1$, $S_2$ and
Wilson loop averages. They parallel relations (\rf{2e14}) of the
2PGIQGFs, but with the additional complication that the iteration
should be carried out simultaneously in $\Phi_{(n+1)}$ and $S_{(n+1)}$.
\par
The final step consists in using expressions (\rf{4e13}) and (\rf{4e14}) 
to bring the right-hand sides of the wave equations (\rf{4e5}), 
(\rf{4e6}), (\rf{4e7}) and (\rf{4e8}) into a form where the wave functions 
appear as acted on by kernels made of Wilson loop averages and their 
derivatives, as well as 2PGIQGFs. The procedure follows similar lines as 
those adopted for the 2PGIQGFs \cite{s1}. Considering in particular the 
wave equation (\rf{4e7}), one finds
\newpage
\bea \lb{4e16}
& &\frac{\bar\delta \Phi_{(1)}(x_1,x_2)}{\bar\delta x_1^{\mu -}}=
\frac{\bar\delta F_{2}(x_2,x_1)}{\bar\delta x_1^{\mu -}}\,
\Phi_{(1)}(x_1,x_2) \nonumber \\
& &\ \ \ \ 
-\frac{\bar\delta^2 F_3(x_2,x_1,y_1)}{\bar\delta x_1^{\mu -}
\bar\delta y_1^{\alpha_1 +}}\Big[\,\Phi_{(1)}(x_1,y_1)\,\gamma^{\alpha_1}
\,S_{2,(2)}(y_1,x_2;x_1)+\,S_1(x_1,y_1)\,\gamma^{\alpha_1}\,
\Phi_{(2)}(y_1,x_2;x_1)\,\Big]
\nonumber \\
& &\ \ \ \ -\Big\{\Big(\frac{\bar\delta \Phi_{(1)}(x_1,y_1)}
{\bar\delta y_1^{\alpha_1 +}}+\Phi_{(1)}(x_1,y_1)\frac{\bar\delta}
{\bar\delta y_1^{\alpha_1 -}}\Big)\,\gamma^{\alpha_1}
\frac{\bar\delta^2 F_{4}(x_2,x_1,y_1,y_2)}{\bar\delta x_1^{\mu -}
\bar\delta y_2^{\alpha_2 +}} \nonumber \\
& &\ \ \ \ \ \ \ \ \ \ \ \ \times S_2(y_1,y_2)\,\gamma^{\alpha_2}\,
S_{2,(3)}(y_2,x_2;x_1,y_1)\nonumber \\
& &\ \ \ \ +\Big(\frac{\bar\delta S_1(x_1,y_1)}
{\bar\delta y_1^{\alpha_1 +}}+S_1(x_1,y_1)\frac{\bar\delta}
{\bar\delta y_1^{\alpha_1 -}}\Big)\,\gamma^{\alpha_1}
\frac{\bar\delta^2 F_{4}(x_2,x_1,y_1,y_2)}{\bar\delta x_1^{\mu -}
\bar\delta y_2^{\alpha_2 +}} \nonumber \\
& &\ \ \ \ \ \ \ \ \ \ 
\times \Big[\,\Phi_{(1)}(y_1,y_2)\,\gamma^{\alpha_2}\,
S_{2,(3)}(y_2,x_2;x_1,y_1)+S_1(y_1,y_2)\,\gamma^{\alpha_2}\,
\Phi_{(3)}(y_2,x_2;x_1,y_1)\,\Big]\Big\}
\nonumber \\
& &\  -\sum_{j=4}^{\infty}\sum_{r=1}^{j-2}
\Big(\frac{\bar\delta S_1(x_1,y_1)}{\bar\delta y_1^{\alpha_1 +}}
+S_1(x_1,y_1)\frac{\bar\delta}{\bar\delta y_1^{\alpha_1 -}}\Big)\,
\gamma^{\alpha_1}\times\cdots \nonumber \\
& &\ \ \    
\times\,\Big(\frac{\bar\delta S_1(y_{r-2},y_{r-1})}
{\bar\delta y_{r-1}^{\alpha_{r-1} +}}+S_1(y_{r-2},y_{r-1})
\frac{\bar\delta}{\bar\delta y_{r-1}^{\alpha_{r-1} -}}\Big)
\gamma^{\alpha_{r-1}} \nonumber \\
& &\ \ \  \times\,\Big(\frac{\bar\delta \Phi_{(1)}(y_{r-1},y_r)}
{\bar\delta y_r^{\alpha_r +}}+\Phi_{(1)}(y_{r-1},y_r)
\frac{\bar\delta}{\bar\delta y_r^{\alpha_r -}}\Big)
\gamma^{\alpha_r}\,
\Big(\frac{\bar\delta S_2(y_r,y_{r+1})}
{\bar\delta y_{r+1}^{\alpha_{r+1} +}}+S_2(y_r,y_{r+1})
\frac{\bar\delta}{\bar\delta y_{r+1}^{\alpha_{r+1} -}}\Big)
\nonumber \\
& &\ \ \  \times\,\gamma^{\alpha_{r+1}}\,\cdots 
\Big(\frac{\bar\delta S_2(y_{j-3},y_{j-2})}
{\bar\delta y_{j-2}^{\alpha_{j-2} +}}+S_2(y_{j-3},y_{j-2})
\frac{\bar\delta}{\bar\delta y_{j-2}^{\alpha_{j-2} -}}\Big)
\gamma^{\alpha_{j-2}}\nonumber \\
& &\ \ \ \times
\frac{\bar\delta^2 F_{j+1}(x_2,x_1,y_1,\ldots,y_{j-1})}
{\bar\delta x_1^{\mu -}\bar\delta y_{j-1}^{\alpha_{j-1} +}}
S_2(y_{j-2},y_{j-1})\,\gamma^{\alpha_{j-1}}\,
S_{2,(j)}(y_{j-1},x_2;x_1,y_1,\ldots,y_{j-2})
\nonumber \\
& &\ -\sum_{j=4}^{\infty}
\Big(\frac{\bar\delta S_1(x_1,y_1)}{\bar\delta y_1^{\alpha_1 +}}
+S_1(x_1,y_1)\frac{\bar\delta}{\bar\delta y_1^{\alpha_1 -}}\Big)\,
\gamma^{\alpha_1}\times\cdots \nonumber \\
& &\ \ \ \ \ \times\,\Big(\frac{\bar\delta S_1(y_{j-3},y_{j-2})}
{\bar\delta y_{j-2}^{\alpha_{j-2} +}}
+S_1(y_{j-3},y_{j-2})\frac{\bar\delta}
{\bar\delta y_{j-2}^{\alpha_{j-2} -}}\Big)\,
\gamma^{\alpha_{j-2}}\nonumber \\
& &\ \ \times
\frac{\bar\delta^2 F_{j+1}(x_2,x_1,y_1,\ldots,y_{j-1})}
{\bar\delta x_1^{\mu -}\bar\delta y_{j-1}^{\alpha_{j-1} +}}\,
\Big[\,\Phi_{(1)}(y_{j-2},y_{j-1})\,\gamma^{\alpha_{j-1}}\,
S_{2,(j)}(y_{j-1},x_2;x_1,y_1,\ldots,y_{j-2}) \nonumber \\
& &\ \ \ \ \ \ \ +S_1(y_{j-2},y_{j-1})\,\gamma^{\alpha_{j-1}}\,
\Phi_{(j)}(y_{j-1},x_2;x_1,y_1,\ldots,y_{j-2})\,\Big]. 
\eea
In the double sum with respect to $r$ and $j$, $y_0$ means $x_1$.
When $r=1$, factors with $S_1$ and derivatives do not exist on the 
left of $\Phi_{(1)}$; when $r=j-2$, factors with $S_2$ and derivatives
do not exist on the right of $\Phi_{(1)}$, except the one preceding
$S_{2(j)}$. 
\par
The kernels appearing in the right-hand side of Eq. (\rf{4e16}) can
be expressed through functional relationships in terms of those
of the 2PGIQGFs. Going back to Eq. (\rf{2e17}) and to the definition 
of the operator $\Sigma_{1n}[S]$ there, to which we assign now a new 
index, 1 or 2, according to its content of quark 1 or quark 2, the wave 
equation (\rf{4e7}) can be rewritten in the form
\be \lb{4e17}
(i\gamma.\partial_1-m_1)\Phi_{(1)}(x_1,x_2)=
\sum_{n=1}^{\infty}\,\Big(\Sigma_{1,1n}[S_1]*\Phi_{(n)}\Big)
+\sum_{n=2}^{\infty}\left(\Big(\frac{\delta\Sigma_{1,1n}[S_1]}{\delta S_1}
**\Phi_{(1)}\Big)*S_{2,(n)}\right),
\ee
where the double-star operator means that after functionally taking the
derivative of $\Sigma_{1,1n}[S_1]$ with respect to $S_1$, the latter is 
replaced in the same place by $\Phi_{(1)}$ and all $S_1$s on the right of 
$\Phi_{(1)}$ are replaced by $S_2$.
\par
A similar expression can also be derived for the right-hand side of
Eq. (\rf{4e8}), using either of the forms (\rf{2e18}) or (\rf{2e19}). 
\par
Equation (\rf{4e16}) parallels Eq. (\rf{2e15}) of the 2PGIQGF. 
High-index wave functions and 2PGIQGFs should in
principle be eliminated in terms of the lowest-index ones using
the functional relations (\rf{4e13}), (\rf{4e14}) and (\rf{2e14}).
Derivative terms of $\Phi_{(1)}$ that appear in the right-hand 
side of Eq. (\rf{4e16}) in high-order terms could 
be eliminated using the first terms of the expansions through an 
iterative procedure. Expansion (\rf{4e16}) should be completed 
by bringing all derivative terms to the right; the result is
very similar to what is obtained in the 2PGIQGF case. Connected parts
of the Wilson loop averages of high-order terms involve an 
increasing number of derivatives, each on a different segment of 
the corresponding polygonal contours \cite{s1}.    
\par
In perturbation theory, each derivative of a Wilson loop 
introduces one multiplicative power of the coupling constant. 
Therefore, for the perturbative regime, terms with the smallest 
number of derivatives would be the dominant ones. For large distances, 
Wilson loop averages are expected to be dominated by minimal 
surfaces \cite{w,mm,bw}. Here also, the dominant terms are those 
with the smallest number of derivatives. This suggests that the 
expansion (\rf{4e16}) has, at least formally, a 
perturbative structure, the decrease in magnitude of the terms being 
estimated by the global number of the derivative operators.
The first term of the above expansions involves a Wilson loop with
one derivative, which, however, vanishes for symmetry reasons. 
The first leading term of the expansion is then the two-derivative 
term, involving a Wilson loop along a triangular contour.
In two-dimensional QCD in the large-$N_c$ limit, this term is 
actually the only one that survives in the above expansions
and brings an indirect confirmation to the previous analysis.
\par
Our remark [in the paragraph following Eq. (\rf{2e19})] concerning 
the interpretation of Eq. (\rf{2e17}) and its analogues also applies
to Eq. (\rf{4e17}) and its analogues. Once the rigid path derivative 
of the wave functions has been evaluated by means of Eq. (\rf{4e16})
or similar ones, then the wave functions have to be considered as 
ordinary complex functions of their arguments $x_1$, $x_2$, $t_1$,
$\ldots$, $t_n$, $P$, etc. The wave equations of the type of 
(\rf{4e17}) become integro\-differential equations.
\par
For completeness, the normalization condition of the wave functions 
is presented in the appendix.
\par 

\section{Spectral representation } \lb{s5}
\setcounter{equation}{0}

A two-particle wave function of colorless fields is a vertex 
function and satisfies, on the basis of the properties of
causality and positivity of physical particle energies, an 
integral representation known as the Deser-Gilbert-Sudarshan 
(DGS) representation \cite{dgs}.
One might try to apply the DGS analysis to the presently defined
wave functions (\rf{4e1}) and (\rf{4e3}). However, an immediate
difficulty arises from the fact that intermediate states needed
for the determination of their singularities are necessarily colored 
objects. Intermediate states, placed inside the operators that are
present in the matrix elements (\rf{4e1}) or (\rf{4e3}), should have 
the quantum numbers of operators made of one quark (belonging to the
defining fundamental representation of the color gauge group) and a 
certain number of gluons (belonging to the adjoint representation); 
these combinations cannot produce color singlet operators. Therefore, 
hadronic intermediate states, which are expected to form the only 
physical states of the theory, do not contribute to the formation 
of the singularities of the wave functions. One then is tempted to 
conclude that the wave functions are free of singularities and are 
entire functions. However, the equations satisfied by the GIQGFs do 
display momentum-space singularities generated by the free quark 
propagators present in them. The same difficulty also appeared in 
the 2PGIQGF case. 
\par
To explain the presence of singularities in the Green's functions 
it is necessary to admit that completeness sums may be considered 
with colored quark and gluon states, irrespective of the fact that 
the latter may not be observable as asymptotic free states.
(A similar conclusion was also drawn in Ref. \cite{oz} concerning
ordinary propagators.) 
It is the solutions of the corresponding equations which should 
ultimately determine their precise properties. 
We further assume the usual spectral and causality properties 
of quantum field theory. The presence of gluon fields, treated here 
covariantly, might introduce in addition negative norm states (but 
still carrying positive energies) in the completeness sum, whose 
main effect could be the change of sign within certain intervals 
of the concerned spectral function \cite{alklvs,fisc,oz}. Therefore, 
no positivity conditions should be imposed in advance on the spectral 
functions (this concerns mainly the 2PGIQGFs). 
\par
More generally, the fact that the intermediate states are colored
objects puts constraints on their specific properties: the colored 
states, when placed inside a gauge invariant operator, separate
the latter into two gauge covariant operators; their contribution 
should reproduce at the end gauge invariant quantities. The study 
of the mechanism of that operation deserves attention in future 
investigations.
\par 
The above general hypotheses were applied to the 2PGIQGF case and 
led to a generalized K\"all\'en-Lehmann representation \cite{kl,lh,s1}. 
The analytic resolution of the 2PGIQGF equation in two-dimensional 
QCD in the large-$N_c$ limit has confirmed the previous hypotheses: 
the quarks and gluons do contribute to the spectral functions 
with positive energies, with the difference that their singularities 
are no longer represented by simple poles, but by an infinite number 
of branch points with a stronger power than poles \cite{s2}.
Furthermore, Lehmann's positivity conditions (or inequalities)
\cite{lh} remain satisfied.   
\par
It is therefore reasonable to continue to apply the same approach
to the study of the spectral representation of the wave functions.
Considering the wave function $\Phi_{(1)}$ [Eq. (\rf{4e3})], one 
can repeat the same analysis as in Ref. \cite{s1} for the 2PGIQGF.
The presence of the gluon field phase factor introduces an infinite
series of additional powers of the denominator of the dispersive 
integral. Defining total and relative coordinates as 
$X=\frac{1}{2}(x_1+x_2)$ and $x=x_1-x_2$, one can factorize the 
plane wave part of the wave function and introduce the internal
wave function as
\be \lb{5e1}
\Phi_{(1)}(P,x_1,x_2)=e^{{\displaystyle -iP.X}}\phi_{(1)}(P,x).
\ee
\par
$\phi_{(1)}(P,x)$ satisfies a generalized form of the DGS 
representation, which we write here, for simplicity, for the total
spin-0 case, ignoring the spinorial content:
\be \lb{5e2}
\phi_{(1)}(P,x)=\sum_{n=1}^{\infty}i\int\frac{d^4k}{(2\pi)^4}
e^{{\displaystyle -ik.x}}\int_{-1/2}^{1/2}\frac{d\beta}{2\pi}
\int_0^{\infty}ds'\,\frac{\,H_{(1),n}(s',\beta)\,
e^{{\displaystyle -i\beta P.x}}}{(k^2-s'+i\varepsilon)^n}.
\ee
The lower bound of the spectral variable $s'$ may actually depend 
on the quark masses, on $P^2$ and on $\beta$. The above general form 
might still be simplified through integrations by parts and 
recombinations. In particular, the sum might lead to a global fractional 
power. In two-dimensional QCD, the resulting power of the denominator 
for the 2PGIQGF case was found 3/2 \cite{s2}.
\par
Representation (\rf{5e2}), or a simpler version of it, could be used 
for the search for solutions of the wave equations. 
According to the solutions that are found, its detailed properties
could be better specified. Nevertheless, we do not dispose of much
theoretical freedom for qualitative changes of representation (\rf{5e2})
without altering some basic property of quantum field theory. That
question might still be reconsidered in the light of the confinement
mechanism that would be found in four-dimensional QCD.  
\par

\section{Chiral symmetry breaking} \lb{s6}
\setcounter{equation}{0}

In the limit of $N_f\times N_c$ massless quarks, it is expected that 
the chiral $SU(N_f)\times SU(N_f)$ symmetry group of QCD undergoes
a spontaneous breakdown to its diagonal subgroup $SU(N_f)_V^{}$ 
\cite{th3,fsby,cg,cw}. On the other hand, the 
Nambu--Jona-Lasinio model \cite{njl} underlines the intimate 
relationship that exists between chiral symmetry breaking and 
the dynamical mass generation of fermions. Baker, Johnson and Lee
considered the case of QED and showed that a similar relationship
also exists there: the generation of a nonzero mass term in the
fermion self-energy in the massless limit entails the existence of 
a zero-mass pseudoscalar bound state solution of the Bethe-Salpeter
equation \cite{bjl}. They argued, however, that such a solution
might not necessarily correspond to an observable boson. The latter
phenomenon is actually a consequence of the axial anomaly problem in
Abelian sectors \cite{aab,bj} and should be avoided by considering 
nonsinglet sectors of a non-Abelian chiral group. 
\par
In this section we aim at showing that the relationship between
dynamical mass generation of fermions and chiral symmetry breaking
also exists in QCD. That question is not new and in the past many
works, based on approaches using the axial-vector Ward-Takahashi
identities, Dyson-Schwinger equations and instantaneous  
approximations of the Bethe-Salpeter kernel with confining 
interactions, have shown the possible validity of such a mechanism
\cite{pg,fgmw,alyopro,ad,aa,lg,brlvs,mrt,alklvs,fisc}. Our proof below 
is more formal and independent of any approximation of the interaction 
kernels.
\par
To this end, we consider the case of massless 
quarks 1 and 2; in this limit the labeling of the 2PGIQGFs with
the quark indices 1 and 2 becomes irrelevant and we may discard
them from our notations. The 2PGIQGF $S_{(1)}$ is decomposed into 
vector and scalar parts \cite{s1}. The part that is most sensitive
to chiral symmetry breaking is the scalar one. We isolate it by
taking the anticommutator of $S_{(1)}$ with the $\gamma_5$ matrix.
It satisfies the equations
\bea 
\lb{6e1}
& &i\gamma.\partial_1[\gamma_5,S_{(1)}(x_1,x_2)]_+^{}=
i\gamma^{\mu}\,\frac{\bar\delta}{\bar\delta x_1^{\mu -}}
[\gamma_5,S_{(1)}(x_1,x_2)]_+^{},\\
\lb
{6e2}
& &[\gamma_5,S_{(1)}(x_1,x_2)]_+^{}(-i\gamma.\stackrel{\leftarrow}
{\partial}_2)=-i\frac{\bar\delta}{\bar\delta x_2^{\mu +}}
[\gamma_5,S_{(1)}(x_1,x_2)]_+^{}\,\gamma^{\mu},
\eea
where $[,]_+^{}$ represents the anticommutator.
\par
The right-hand sides of Eqs. (\rf{6e1}) and (\rf{6e2}) are obtained 
from Eq. (\rf{2e15}) and its adjoint with respect to $x_2$ and more 
schematically from Eqs. (\rf{2e17})-(\rf{2e19}). We have for 
the rigid path derivative with $x_1$,
\bea \lb{6e3}
& &\frac{\bar\delta}{\bar\delta x_1^{\mu -}}
[\gamma_5,S_{(1)}(x_1,x_2)]_+^{}=
\frac{\bar\delta F_{2}(x_2,x_1)}{\bar\delta x_1^{\mu -}}\,
[\gamma_5,S_{(1)}(x_1,x_2)]_+^{} \nonumber \\
& &\ \ \ \ 
-\frac{\bar\delta^2 F_3(x_2,x_1,y_1)}{\bar\delta x_1^{\mu -}
\bar\delta y_1^{\alpha_1 +}}\Big\{
[\gamma_5,S_{(1)}(x_1,y_1)]_+^{}\,\,\gamma^{\alpha_1}
\,S_{2,(2)}(y_1,x_2;x_1)\nonumber \\
& &\ \ \ \ \ \ +S_1(x_1,y_1)\,\gamma^{\alpha_1}\,
[\gamma_5,S_{(2)}(y_1,x_2;x_1)]_+^{}\,\Big\}\,+\ldots\,,
\eea
where the dots correspond to similar terms coming from the 
contribution of the sum contained in Eq. (\rf{2e15}).
In schematic form, the equation relative to $x_1$ is 
\be \lb{6e4}
i\gamma.\partial_1[\gamma_5,S_{(1)}]_+^{}=
\sum_{n=1}^{\infty}\,\Big(\Sigma_{1n}[S_{(1)}]*
[\gamma_5,S_{(n)}]_+^{}\Big)
+\sum_{n=2}^{\infty}\left(\Big(\frac{\delta\Sigma_{1n}[S_{(1)}]}
{\delta S_{(1)}}**[\gamma_5,S_{(1)}]_+^{}\Big)*S_{(n)}\right),
\ee
where $\Sigma_{1n}$ and the star operation were introduced in Eq.
(\rf{2e17}), while the definition of the double-star operator is the 
same as in Eq. (\rf{4e17}), with the only difference that the quark 
indices, 1 or 2, are now removed from the Green's functions.
\par
Comparing the above equation with Eq. (\rf{4e17}) (in the limits 
$m_1=m_2=0$ and removing the quark indices 1 or 2), one 
immediately concludes that $[\gamma_5,S_{(1)}(x_1,x_2)]_+^{}$ satisfies 
the same equation as $\Phi_{(1)}(x_1,x_2)$, provided the following 
correspondences are also done: $[\gamma_5,S_{(n)}]_+^{}$
$\rightarrow$ $\Phi_{(n)}$, $n=1,2,\ldots,\infty$, the general case 
of $n$ being itself established from the equations of $S_{(n)}$.
A similar conclusion also holds for the adjoint equations corresponding
to $x_2$. On the other hand, the 2PGIQGFs do not depend on the bound 
state total momentum vector $P$. Therefore, the above correspondence is 
possible only if the $\Phi$s are independent of $P$; this is possible 
if $P=0$ in the corresponding bound state, which in turn implies $P^2=0$.
\par
We thus arrive at the conclusion that if the 2PGIQGF $S_{(1)}$ has,
in the massless quark limit, a nontrivial normalizable scalar part,
then the latter, multiplied by the $\gamma_5$ matrix, is a solution
of the bound state equation with zero mass, in the limit of
zero total momentum. This is an indication of the existence of
a pseudoscalar Goldstone boson in the bound state spectrum. The 
complete expression of the corresponding wave function for nonzero
$P$ should be searched for from the bound state equations themselves;
that part is not given by the 2PGIQGFs.
\par
The previous exact relationship between the scalar part of the 2PGIQGF
$S_{(1)}$ and the wave function with zero total momentum could also be 
stated in any truncation scheme adopted as an approximation for the 
resolution of the corresponding equations. Equation (\rf{6e4}), when 
written explicitly, has a structure similar to that of Eq. (\rf{4e16}). 
Therefore, any truncation scheme in one of the equations has its equivalent 
truncation scheme in the other equation. This is easily checked with 
Eq. (\rf{6e3}), where dropping the dots would amount to truncating 
the series beyond two derivative terms in the kernels. One checks in 
Eq. (\rf{4e16}) that the same approximation, keeping in the right-hand
side the first three terms, reproduces the structure of Eq. (\rf{6e3}). 
This property remains true order by order of the expansion based on the 
number of derivatives. It allows, on practical grounds, the use of 
approximations that remain compatible with chiral symmetry.
\par  
It is worth noting that the scalar part of the 2PGIQGF $S_{(1)}$ does
not necessarily correspond to a conventional mass term which would give 
rise to a pole structure in the quark Green's function. Rather, the 
property that quarks are confined suggests that it would possess a more 
complicated structure. In two-dimensional QCD, the resolution of the 
problem led to the appearance of an infinite set of branch points, at 
dynamically generated mass values $M_1$, $M_2$, $\ldots$, $M_n$, $\ldots$, 
with stronger singularities than simple poles \cite{s2}.  
\par      
The above criterion for chiral symmetry breaking is only a sufficient
one. This is a consequence of the fact that the equations satisfied 
by $\Phi$ are linear in $\Phi$ (given the expressions of the $S$s),
while the equations satisfied by $[\gamma_5,S]_+^{}$ are nonlinear,
since the latter are also contained in the $S$s present in the equations.
\par

\section{Summary and concluding remarks} \lb{s7}
\setcounter{equation}{0}

Phase factors along polygonal lines allow a simple classification
of gauge invariant quark Green's functions according to the number
of segments they contain on the lines and permit a systematic 
investigation of their properties through their equations of motion. 
The latter can be reexpressed as integro\-diffe\-rential equations in 
which the kernels are essentially represented by Wilson loop averages 
along polygonal contours with an increasing number of sides and derivatives. 
\par
This approach was applied to the cases of two-point and four-point
Green's functions in QCD, leading in the latter case to bound state 
equations for quark-antiquark systems. The functional 
relationships between the kernels of the bound state equations and 
those of the two-point functions were displayed. A sufficient criterion 
for spontaneous chiral symmetry breaking was derived, relating the 
Goldstone boson wave function in the zero total momentum limit with 
the scalar part of the two-point Green's function.   
\par
The idea of relating the perturbative degree of a kernel with the 
number of sides and derivatives of the Wilson loop contours offers
promising perspectives for practical applications of the equations
obtained thus far. In this case, the dominant terms of the interaction
kernels would come from the simplest contours and the least number
of derivatives. That feature is also manifest in two-dimensional
QCD in the large-$N_c$ limit. An analysis of the structure of the 
present equations in two dimensions would provide us with a 
simplified framework for the understanding of various mechanisms at
work and this in turn might serve as helpful guidance for future
resolutions of the relevant problems in four dimensions.
\par

%\newpage
\vspace{0.5 cm}
\noindent
\textbf{{\large Acknowledgements}}
\par
\vspace{0.25 cm}
I thank Wolfgang Lucha for providing me references concerning 
approaches to chiral symmetry breaking. This work is partially 
supported by the EU I3HP project ``Study of Strongly Interacting 
Matter'' (acronym HadronPhysics3, Grant Agreement No. 283286). 
\par
\newpage 

\appendix
\renewcommand{\theequation}{\Alph{section}.\arabic{equation}}

\section{Normalization condition} \lb{ap2}
\setcounter{equation}{0}

The normalization condition of the wave functions is usually 
obtained by acting on the four-point Green's function with its 
inverse. In the present case, the inhomogeneous parts of the
equations of motion (\rf{3e3}) and (\rf{3e4}) being two-point
Green's functions, one needs to use, for $G_{(1)}$, simultaneously
both equations. One obtains
\bea \lb{be1}
& &(i\gamma.\partial_1-m_1-i\gamma^{\mu}
\frac{\bar\delta}{\bar\delta x_1^{\mu -}})\,G_{(1)}(x_1,x_2;x_2',x_1')
\,(-i\gamma.\stackrel{\leftarrow}{\partial}_2-m_2-
i\gamma^{\nu}
\frac{\stackrel{\leftarrow}{\bar\delta}}{\bar\delta x_2^{\nu +}})
\nonumber \\
& &\ \ \ \ \ \ \ \ \ \ \ 
=i^2\delta^4(x_1-x_1')\delta^4(x_2-x_2').
\eea
[$F_2(x_2,x_1)=1.$] 
\par
After calculating the effects of the rigid path derivatives,
one passes to total momentum space $P$ by taking the Fourier
transform with respect to the total coordinate difference
$X-X'=\frac{1}{2}(x_1+x_2)-\frac{1}{2}(x_1'+x_2')$, while remaining
in the relative coordinate spaces $x=(x_1-x_2)$ and $x'=(x_1'-x_2')$. 
Designating by $L(P,x)$ the operator acting in the left-hand side of 
Eq. (\rf{be1}), the latter can be written in condensed form as
\be \lb{be2}
L(P,x)*G_{(1)}(P,x,x')=i^2\delta^4(x-x'),
\ee
where the star notation represents here the complete series of terms
obtained by the action of $L(P,x)$ on $G_{(1)}$ as in Eqs. (\rf{2e15})
and (\rf{4e16}), which involves linearly the set of Green's 
functions $G_{(n)}$ ($n=1,2,\ldots$).
\par
We specify with the label $k$ one of the bound states of the system
and isolate its contribution and that of its antiparticle in $G_{(1)}$ 
from the rest,
\be \lb{be3}
G_{(1)}(P,x,x')=\frac{i\phi_{(1)k}(P_k,x)\overline{\phi}_{(1)k}(P_k,x')}
{P^2-P_k^2+i\varepsilon}+iR_{(1)k}(P,x,x'),
\ee
where the $\phi$s represent the internal part of the wave functions,
as defined in Eq. (\rf{5e1}). (The chronological product being evaluated
with the time components of the $x$s, we have also 
$\mathbf{P_k}=\mathbf{P}$.) Decomposition (\rf{be3}) is replaced in
Eq. (\rf{be2}) and then both sides of the latter are multiplied with
$\overline{\phi}_{(1)k}(x)$ and integrated with respect to $x$, with
the trace on the spinor indices taken. Taking into account the fact 
that the operator $L(P_k)$ annihilates $\phi_{(1)k}$, one obtains
\bea \lb{be4}
& &\overline{\phi}_{(1)k}(x)\Big(\frac{L(P,x)-L(P_k,x)}
{P^2-P_k^2+i\varepsilon}\Big)*\phi_{(1)k}(x)\,\overline\phi_{(1)k}(x')
+\overline{\phi}_{(1)k}(x)L(P,x)*R_{(1)k}(P,x,x')\nonumber \\
& &\ \ \ \ \ \ \ \ \ \ =i\overline{\phi}_{(1)k}(x').
\eea
\par
At this stage, when working with conventional Green's functions,
one takes the limit $P^2\rightarrow P_k^2$. The convolutive 
nature of $L(P)$ then allows it also to act on the left and the
remainder contribution disappears from the equation. In the present
case, the operator $L(P)$ acts specifically on the right and without 
further information it is not entitled to be converted to the left. 
\par
To go further, we consider, as in Eq. (\rf{be1}), equations of motion
of $G_{(1)}$ concerning its two right arguments 
and define correspondingly an operator $\stackrel{\leftarrow}{L}(P)$ 
acting from the right,
\be \lb{be5}
G_{(1)}(P,x'',x)*\stackrel{\leftarrow}{L}(P,x)=i^2\delta^4(x''-x).
\ee
From Eqs. (\rf{be2}) and (\rf{be5}) one deduces
\bea \lb{be6}
& &G_{(1)}(P,x'',x)L(P,x)*G_{(1)}(P,x,x')=
G_{(1)}(P,x'',x)*\stackrel{\leftarrow}{L}(P,x)G_{(1)}(P,x,x')
\nonumber \\
& &\ \ \ \ \ \ \ \ \ \ =i^2G_{(1)}(P,x'',x'),
\eea 
which entails a weak form of conversion of the operator $L(P)$ from
right to left. Since this result is true for any $P$, $x''$ and $x'$,
one might adopt the assumption that it remains true also for parts
of $G_{(1)}$. Adopting the latter assumption, and taking the limit 
$P^2\rightarrow P_k^2$ in Eq. (\rf{be4}), one obtains the two 
relations,  
\bea 
\lb{be7}
& &\frac{1}{i}\int d^4x\,\overline{\phi}_{(1)k}(x)
\frac{\partial L(P,x)}{\partial P^2}\Big|_{P^2=P_k^2}^{}*\phi_{(1)k}(x)=1,
\\
\lb{be8}
& &\int d^4x\,\overline{\phi}_{(1)k}(x)L(P_k,x)*R_{(1)k}(P_k,x,x')=0,
\eea  
which display the normalization and orthogonality conditions.
In Eq. (\rf{be7}), the derivative of $L(P,x)$ with respect to $P^2$
is understood in the sense that one first evaluates the effect of 
$L(P)$ on $\phi_{(1)k}$ and then takes in the resulting kernels the
corresponding derivative. 
\par
Isolating in $R_{(1)k}$ the contribution of another bound state,
specified by a label $m$, and assuming its independence from the 
rest, one obtains the more precise orthogonality relation
\be \lb{be9}
\int d^4x\,\overline{\phi}_{(1)k}(x)\Big(\frac{L(P_k,x)-L(P_m,x)}
{P_k^2-P_m^2}\Big)*\phi_{(1)m}(x)=0,\ \ \ \ \ \ m\neq k.
\ee
\par
We notice that the free part of the normalization kernel in Eq.
(\rf{be7}), coming from the two Dirac operators, is the same as 
for the Bethe-Salpeter wave functions \cite{nk}.
In the nonrelativistic limit, decomposing $\phi$ into $2\times 2$ 
components, those which are dominant satisfy the properties
$\gamma_0\phi=-\phi\gamma_0=\phi$.   
\par

\end{document}